# Optical bistability and cooling of a mechanical oscillator induced by radiation pressure in a hybrid optomechanical system

Bijita Sarma and Amarendra K. Sarma[*]
*Department of Physics, Indian Institute of Technology Guwahati, Guwahati-781039, Assam, India*
*[*]aksarma@iitg.ernet.in*

We investigate theoretically the effect of optical feedback from a cavity containing an ultracold two level atomic ensemble, on the bistable behavior shown by mean intracavity optical field and the ground state cooling effect of the mechanical oscillator in an optomechanical cavity resonator. The optical bistability can be controlled by tuning the frequency and power of the single driving laser as well as by varying the atom-cavity coupling strength in the atomic cavity. Study of the cooling of the mechanical oscillator, in both good and bad cavity limits, reveals that the hybrid system is more efficient in cooling in comparison to a generic optomechanical setup, even at room temperature. In essence, our work emphasizes the impact of the coupling with the atomic cavity on the radiation pressure effects in the optomechanical cavity.

## I. INTRODUCTION

A photon scattered from an object transfers momentum to the scatterer, thereby applying radiation pressure force on it. Braginsky and his co-workers in their seminal papers [1-3], predicted long ago that the radiation pressure induced by the optical field confined in a cavity resonator can couple the optical and mechanical modes of the cavity. If we consider an optomechanical cavity having a movable end mirror, driven by a strong laser pump, the radiation pressure force applied by the cavity optical field becomes influential enough to set even macroscopic end mirrors into motion. The motion of the mechanical oscillator modulates the length of the cavity and the optical intensity in the cavity gets altered in turn. This type of system shows high nonlinearity between optical field and mechanical motion, acting analogous to a Kerr-medium [4-5]. In recent years, optomechanical systems have drawn tremendous research interest owing to the possibility of implementing these systems in ground state cooling of mesoscopic mechanical oscillator [6-9], continuous variable entanglement of optical and mechanical modes [10-12], optomechanically induced transparency [13-16], nonclassical state generation [17-19] and quantum state transfer between different modes [20-23], among others.

Currently hybrid optomechanical systems [24-30] are highly in focus due to the versatility of both optical and mechanical components in coupling to different systems such as spins, cold atoms, superconducting qubits etc. In this work, we explore the effects of radiation pressure force in a hybrid optomechanical system consisting of two cavities, one optomechanical and the other containing an ultracold two level atomic ensemble, coupled by a single pump laser. In particular, we study the bistable behavior shown by the cavity optical field in the optomechanical cavity with and without feedback from the atomic cavity. Optical bistability inside a cavity with finite decay time, arising due to the dynamic backaction induced by radiation pressure has been studied in various optomechanical systems [31-33]. Here, we discuss the controllability of the bistable behavior depending on the system parameters provided by the feedback cavity, allowable under experimental considerations.

Furthermore, we investigate the ground state cooling of the mechanical oscillator which is a prerequisite for observing quantum effects in optomechanical systems. We study cooling of the mechanical oscillator with and without the atomic cavity feedback and compare the results.

This paper is organized as follows. In Section II we describe the total Hamiltonian of the system and derive the quantum Langevin equations for the system operators. Section III is devoted to the analysis of bistable behavior shown by the mean intracavity optical field in the optomechanical cavity. Section IV discusses cooling of the mechanical oscillator followed by conclusion of our work in Section V.

## II. MODEL AND THEORY

We consider a hybrid optomechanical system consisting of two cavities A and C as shown schematically in Fig. 1. Cavity C, with both the end mirrors fixed contains an ensemble of ultracold two level atoms. Cavity A consists of one fixed mirror and another movable mirror with frequency $\omega_m$, effective mass $m$ and decay rate $\gamma_m$. Cavity A is driven by an intense pump laser of frequency $\omega_L$, which exerts a radiation pressure force on the movable end mirror. The output optical field from the cavity A drives the cavity C, and the output from cavity C is again fed back into the cavity A. We consider the two cavities to have decay rates $k_C \gg k_A$, so that the atomic cavity follows the optomechanical cavity adiabatically.

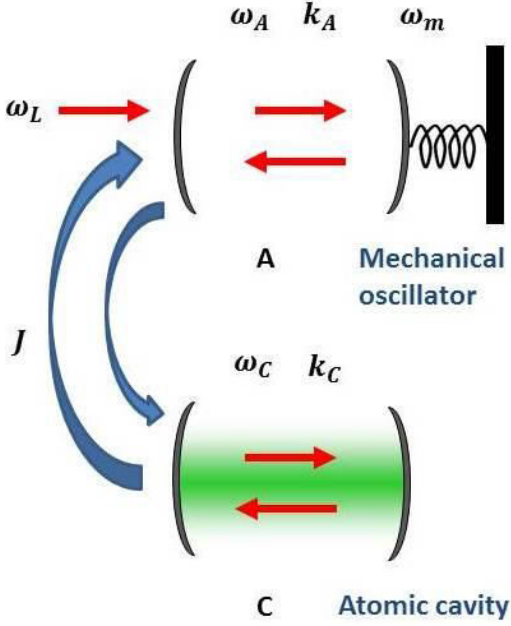

Fig. 1: A hybrid optomechanical cavity setup with an optomechanical cavity A and a feedback cavity C containing ultracold atomic ensemble, coupled optically.

The Hamiltonian of the whole system, in a frame rotating with the driving laser frequency $\omega_L$, is given by

$$H = \hbar\Delta_A a^\dagger a + \hbar\Delta_C c^\dagger c + \frac{p^2}{2m} + \frac{1}{2} m\omega_m^2 q^2 + \frac{1}{2}\hbar\Delta_{at}\sigma_3 + \hbar g_{at}(c^\dagger\sigma_{12} + c\sigma_{21}) - \hbar g_{OM} a^\dagger a q + \hbar J(c^\dagger a + a^\dagger c) + i\hbar\varepsilon_A(a^\dagger - a) + i\hbar\varepsilon_C(c^\dagger - c) \quad (1)$$

Here, the first and second terms represent the energy of the cavity modes in the two cavities A and C respectively. $\Delta_A = \omega_A - \omega_L$ and $\Delta_C = \omega_C - \omega_L$ are the cavity detunings with $\omega_A$ and $\omega_C$ being the corresponding cavity resonance frequencies. The third and fourth terms give the energy of the mechanical oscillator expressed in terms of the position and momentum operators $q$ and $p$ satisfying the commutation relation $[q,p] = i\hbar$. The fifth term is the energy of the two-level atomic ensemble trapped in the cavity C where, $\Delta_{at} = \omega_{at} - \omega_L$, is the detuning of the atomic resonance frequency and $\sigma_3 = \sigma_{22} - \sigma_{11}$, with $\sigma_{22}$ and $\sigma_{11}$ being the atomic populations in the excited and ground levels respectively. The sixth term describes the interaction of the atomic ensemble with the optical field in the cavity C, $g_{at}$ being the mutual coupling constant. The seventh term is the optomechanical interaction term, where $g_{OM} = \frac{\omega_A}{L}$ is the optomechanical coupling constant between the cavity field and the mechanical oscillator in cavity A. The eighth term accounts for the coupling between the two cavities with $J = \sqrt{k_A k_C}$ [29]. The last two terms represent the two cavities driven by the pump with frequency $\omega_L$ and amplitudes $\varepsilon_A = \sqrt{\frac{2k_A P_{in,A}}{\hbar\omega_L}}$ and $\varepsilon_C = \sqrt{\frac{2k_C P_{in,C}}{\hbar\omega_L}}$; $P_{in,A}$ and $P_{in,C}$ being the input powers in the two cavities.

To study the effect of feedback into cavity A, we first need to analyze the cavity field dynamics in cavity C. For this, the Hamiltonian for cavity C in a frame rotating with frequency $\omega_L$ would be

$$H_C = \hbar\Delta_C c^\dagger c + \frac{1}{2}\hbar\Delta_{at}\sigma_3 + \hbar g_{at}(c^\dagger\sigma_{12} + c\sigma_{21}) + i\hbar\varepsilon_C(c^\dagger - c) \quad (2)$$

The time evolution of the system operators are given by nonlinear Heisenberg-Langevin equations

$$\frac{dc}{dt} = -(k_C + i\Delta_C)c - ig_{at}\sigma_{12} + \varepsilon_C + \sqrt{2k_C}c_{in}(t) \quad (3)$$

$$\frac{d\sigma_{12}}{dt} = -(\gamma_{at} + i\Delta_{at})\sigma_{12} - \gamma_{at}\sigma_{12} + \sqrt{2\gamma_{at}}c_{in}(t) \quad (4)$$

where, $\gamma_{at}$ is the atomic coherence decay rate and $c_{in}$ is the input vacuum noise operator with zero mean value and nonzero correlation function given by [34]:

$$\langle c_{in}(t)c_{in}^\dagger(t')\rangle = \delta(t-t') \quad (5)$$

Assuming the system operators under mean field approximation and considering $\langle\sigma_{22}\rangle = 0$ and $\langle\sigma_{11}\rangle = N$, i.e. atoms populating only the ground state, the steady state operators are given by

$$c_S = \frac{\varepsilon_C}{k_C + i\Delta_C + \frac{g_{at}^2 N}{\gamma_{at} + i\Delta_{at}}} \quad (6)$$

$$\sigma_{12,S} = \frac{-ig_{at}c_S N}{\gamma_{at} + i\Delta_{at}} \quad (7)$$

The output of the cavity C is coupled to cavity A. Therefore, we can consider, $\varepsilon_C = iJa_S$. Now, defining two dimensionless position and momentum operators $Q$ and $P$ as $Q = \sqrt{\frac{m\omega_m}{\hbar}}\, q$ and $P = \sqrt{\frac{1}{m\hbar\omega_m}}\, p$ for the mechanical oscillator, the equations of motion for the operators for cavity A are given by

$$\frac{dQ}{dt} = \omega_m P \tag{8}$$

$$\frac{dP}{dt} = \omega_m \chi a^\dagger a - \omega_m Q - \gamma_m P + \xi \tag{9}$$

$$\frac{da}{dt} = -i\left(\Delta_A - \omega_m \chi Q + i\frac{J^2}{k_C + i\Delta_C + \frac{g_{at}^2 N}{\gamma_{at} + i\Delta_{at}}}\right)a - k_A a + \varepsilon_A + \sqrt{2k_A}\,a_{in} \tag{10}$$

where, $\chi = \frac{\omega_A}{\omega_m L}\sqrt{\frac{\hbar}{m\omega_m}}$ is the scaled coupling constant. $a_{in}$ is the input vacuum noise operator for cavity A given by [34]:

$$\langle a_{in}(t) a_{in}^\dagger(t') \rangle = \delta(t - t') \tag{11}$$

$\xi$ is the Brownian noise operator associated with the damping of the mechanical oscillator, with zero mean value and nonzero correlation function [35] given as

$$\langle \xi(t)\xi(t') \rangle = \frac{1}{2\pi}\frac{\gamma_m}{\omega_m}\int \omega e^{-i\omega(t-t')}[1 + \coth(\frac{\hbar\omega}{2k_B T})]\,d\omega \tag{12}$$

## II. OPTICAL BISTABILITY

Bistability is an inevitable behavior observed in nonlinear systems. The inherent nonlinearity in the equations of motion of our system indicates observation of such effects through optomechanical coupling. Considering that mean values of the system operators can be factorized, the steady state solutions of the equations (8)-(10) are given by

$$Q_S = \chi |a_S|^2 \tag{13}$$

$$P_S = 0 \tag{14}$$

$$a_S = \frac{\varepsilon_A}{k_A - \frac{J^2}{k_C + i\Delta_C + \frac{g_{at}^2 N}{\gamma_{at} + i\Delta_{at}}} + i\Delta} \tag{15}$$

where, $\Delta = \Delta_A - \omega_m \chi Q_S$ is the effective detuning in the optomechanical cavity.

Under intense laser pumping, equations (8)-(10) can be linearized by expanding the system operators around their classical mean value and adding zero mean fluctuating operators, as $O = O_S + \delta O$. The quantum Langevin equations for the fluctuation operators are given by

$$\dot{f}(t) = Af(t) + n(t) \tag{16}$$

where, $f^T(t) = (\delta Q, \delta P, \delta X, \delta Y)$ and $n^T(t) = (0, \xi, \sqrt{2k_A}\delta X_{in}, \sqrt{2k_A}\delta Y_{in})$ with $\delta X = \frac{\delta a - \delta a^\dagger}{2}$, $\delta Y = \frac{i(\delta a^\dagger - \delta a)}{2}$, the quadrature operators and $\delta X_{in} = \frac{\delta a_{in} - \delta a_{in}^\dagger}{2}$, $\delta Y_{in} = \frac{i(\delta a_{in}^\dagger - \delta a_{in})}{2}$, the input-noise operators. Considering $a_S$ to be real, the drift matrix of the system is given by

$$A = \begin{pmatrix} 0 & \omega_m & 0 & 0 \\ -\omega_m & -\gamma_m & \sqrt{2}\omega_m \chi a_S & 0 \\ 0 & 0 & E & F \\ \sqrt{2}\omega_m \chi a_S & 0 & -F & E \end{pmatrix}$$

where, for simplicity we have denoted the terms as:

$$E = -k_A + \frac{J^2\{(\gamma_{at}^2 + \Delta_{at}^2)k_C + g_{at}^2 N\gamma_{at}\}}{D},$$

$$F = \Delta_A - \omega_m \chi^2 |a_S|^2 + \frac{J^2\{(\gamma_{at}^2 + \Delta_{at}^2)\Delta_C - g_{at}^2 N\Delta_{at}\}}{D},$$

$$D = (\gamma_{at} k_C - \Delta_{at}\Delta_C + g_{at}^2 N)^2 + (\Delta_{at} k_C + \gamma_{at}\Delta_C)^2 \tag{17}$$

Equations (13)-(15), show that the steady state value of cavity field operator $a$ gives a third order nonlinear equation having three roots. Out of the three roots, only the maximum and minimum roots are stable, the middle one being unstable. The stable roots are obtained only when the all the eigenvalues of matrix A have negative real parts. The stability analysis is done by utilizing the Routh-Hurwitz criterion [36] that gives conditions for stability in our system as:

$$\gamma_m > 2E,$$

$$(\gamma_m - 2E)(-2\gamma_m E + E^2 + F^2 + \omega_m^2) - (\gamma_m E^2 + \gamma_m F^2 - 2E\omega_m^2) > 0,$$

$$\begin{aligned}(\gamma_m - 2E)[(-2\gamma_m E + E^2 + F^2 + \omega_m^2)(\omega_m^2 E^2 + \omega_m^2 F^2 \\ - 2E\omega_m^2) \\ - (\omega_m^2 E^2 + \omega_m^2 F^2 - 2\omega_m^3 \chi^2 a_S^2 F)(\gamma_m \\ - 2E)] - (\gamma_m E^2 + \gamma_m F^2 - 2E\omega_m^2)^2 \\ > 0\end{aligned}$$

$$\omega_m^2 E^2 + \omega_m^2 F^2 - 2\omega_m^3 \chi^2 a_S^2 F > 0 \tag{18}$$

To analyze the bistability behavior, first we consider the case for $J = 0$, i.e. without coupling the atomic cavity. Figure 2(a) shows the behavior of intra-cavity optical intensity in the optomechanical cavity denoted in terms of $\chi Q_S$ with respect to normalized cavity detuning $\Delta_A/\omega_m$. For driving laser power $P = 0.3\,\mu W$, the mean intracavity intensity curve is nearly Lorentzian. With increasing power of the driving laser, it is clearly visible that the bistability occurs for larger cavity detuning.

Figure 2(b) is the hysteresis curve for the mean intracavity intensity vs. varying input power. This curve clearly indicates the bistable behavior of the intracavity photon intensity without the feedback cavity. If we start scanning with a low driving power and gradually increase the power, the intra-cavity intensity initially follows the lower stable branch. When it reaches the first bistable point $P_1$, it jumps to the upper stable branch and continues to follow that branch for further increasing laser power. Now if we start reducing the input laser power the intra-cavity intensity will be found to be decreasing by following the upper branch at first; however whenever it reaches the second bistable point $P_2$, it will jump down to the lower stable branch and continue to decrease further along that branch.

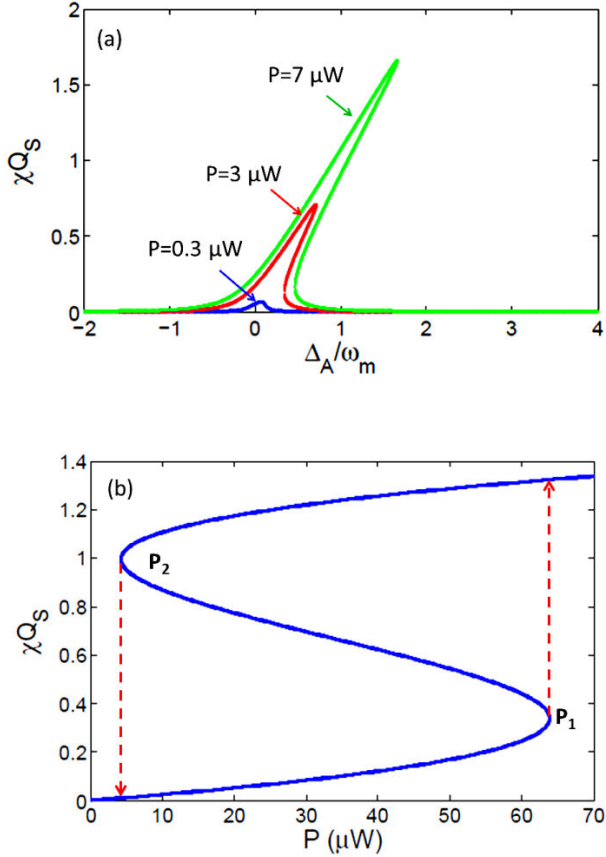

Fig.2: Plot of (a) $\chi Q_S$ vs $\frac{\Delta_A}{\omega_m}$ for $P = 0.3\ \mu W$ (blue solid line), $3\ \mu W$ (red solid line) and $7\mu W$ (green solid line), (b) $\chi Q_S$ vs $P$ for detuning $\Delta_A = \omega_m$. Other system parameters used are $L = 1mm$, $m = 10\ ng$, $\lambda = 794.98\ nm$, $\omega_m = 10\ MHz$, $k_A = 0.1\ \omega_m$, $Q = 10^7$.

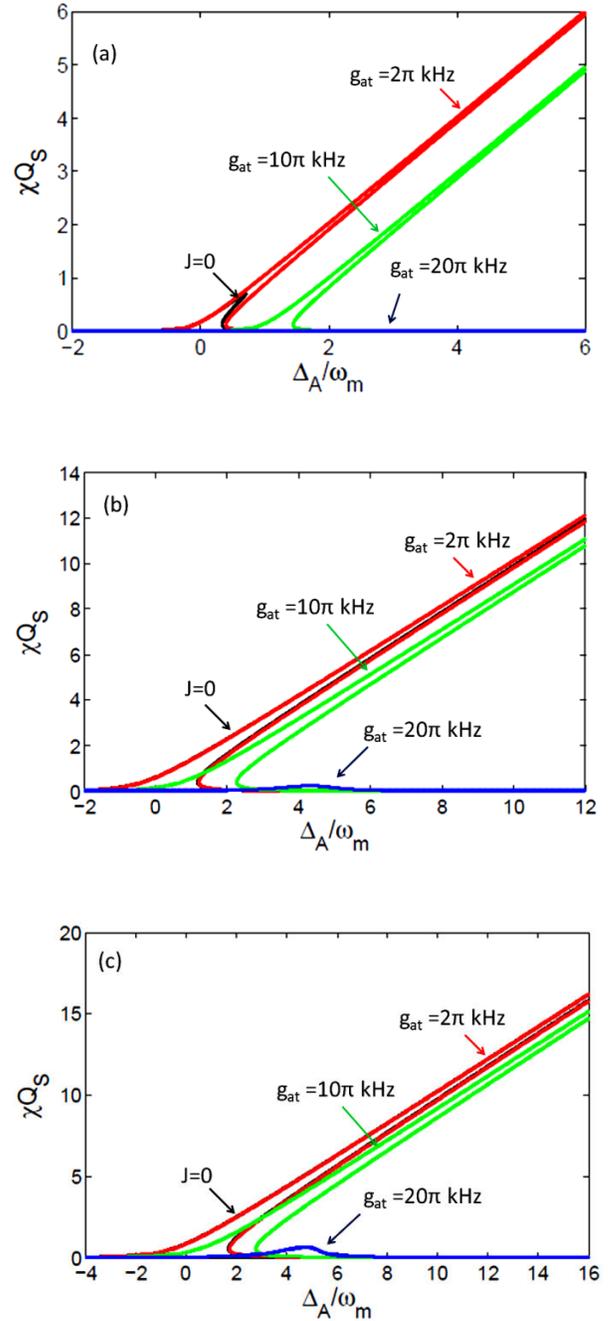

Fig.3: Mean intracavity intensity in terms of $\chi Q_S$ versus the normalized cavity pump detuning $\frac{\Delta_A}{\omega_m}$ in the absence of feedback ($J = 0$) and in presence of feedback with different atom-cavity coupling calculated for different pump powers (a) $P = 3\ \mu W$, (b) $P = 100\ \mu W$, (c) $P = 300\ \mu W$. Parameters used are $k_C = 2\pi \times 21.5\ MHz$,

$N = 10^8$, $\gamma_{at} = 2\pi \times 2.875\, MHz$, $\Delta_{at} = 500\gamma_{at}$, $g_{at} = 2\pi\, kHz$, others same as Fig. 2.

In figures 3(a)-3(c), we have demonstrated the variation of mean intra-cavity intensity in presence of the coupling $J$ between the two cavities. The plots elucidate that the magnitude of the atom-field coupling in cavity C plays significant role in controlling the bistable behavior of optical field in cavity A. We have shown the variations for three different input powers, $P = 3\,\mu W$, $100\,\mu W$ and $300\,\mu W$. With increasing power, for larger value of $g_{at}$, the bistability is seen to occur for larger detuning in the optomechanical cavity.

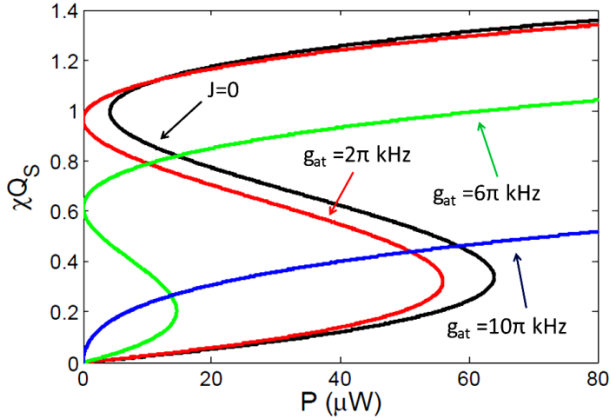

Fig. 4: Plot of $\chi Q_S$ vs $P$, with cavity detuning $\Delta_A = \omega_m$, for $J = 0$ (black solid line) and $J \neq 0$ cases: $g_{at} = 2\pi\, kHz$ (red solid line), $g_{at} = 6\pi\, kHz$ (green solid line) and $g_{at} = 10\pi\, kHz$ (blue solid line).

Figure 4 shows the variation of $\chi Q_S$ vs. $P$ for the uncoupled situation as well as for the feedback coupling with different atom-field coupling strengths. For the cavity detuning in red sideband regime, $\Delta_A = \omega_m$, bistability vanishes for the atom-field coupling $g_{at} = 10\pi\, kHz$. So, the plots clearly indicate that even with higher input power we get stable behavior for larger atom-field coupling. It should be noted that the mean intracavity intensity is highly system specific. Therefore, we can conclude that the extra controlling parameters provided by the feedback from the atomic cavity present us with more flexibility in switching the intra-cavity intensity in cavity A, between the two stable branches.

## II. COOLING OF THE MECHANICAL OSCILLATOR

In the previous section, we studied one of the most important consequences of radiation pressure backaction in the hybrid system. Here, we intend to explore cooling of the mechanical mirror that is another celebrated consequence of radiation pressure.

The radiation pressure force $F_{RP}$ in an optomechanical system is given by $F_{RP} = -\frac{\delta H_{int}}{\delta q}$, where $H_{int}$ is the interaction Hamiltonian of the system. Using this relation, we estimate the radiation pressure force for the system considered in our work, at frequency $\omega_m$ to be:

$$\delta F_{RP}(\omega_m) = \hbar g_{OM} a_S \{\delta a(\omega_m) + \delta a^\dagger(\omega_m)\} \quad (19)$$

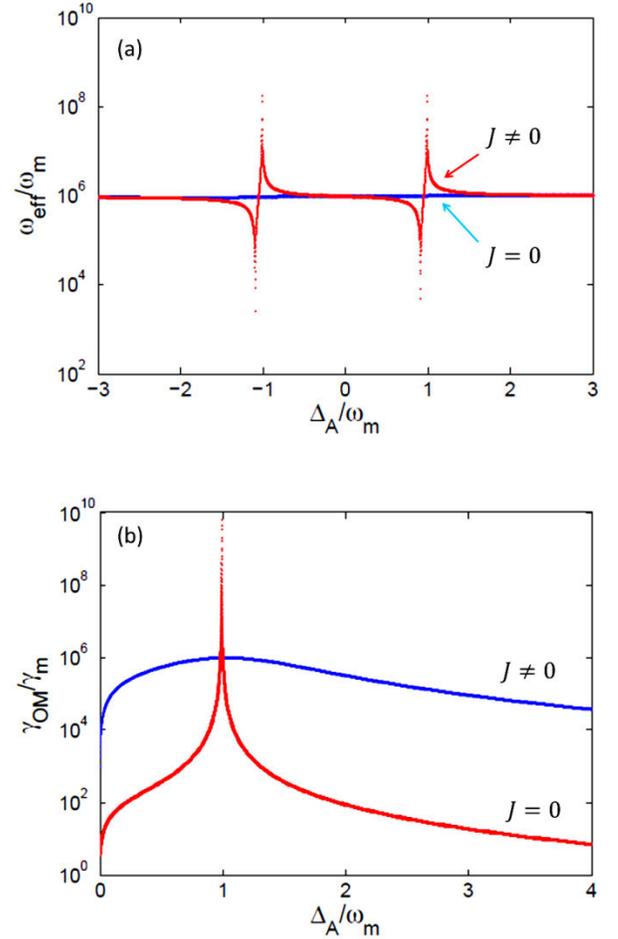

Fig. 5: Plot of (a) $\frac{\omega_{eff}}{\omega_m}$ and (b) $\frac{\gamma_{OM}}{\gamma_m}$ vs $\Delta_A/\omega_m$ in the good cavity limit for $J = 0$ (blue solid line) and $J \neq 0$ (red solid line), for $P = 3\,\mu W$. Other system parameters used are $L = 1\,mm$, $m = 10\,ng$, $\lambda = 794.98\,nm$, $k_C = 2\pi \times 21.5\,MHz$, $k_A = 2\pi \times 250\,kHz$, $\omega_m = 2\pi \times 350\,kHz$, $Q = 10^7$, $N = 10^8$, $\gamma_{at} = 2\pi \times 2.875\,MHz$, $\Delta_{at} = 500\gamma_{at}$, $g_{at} = 2\pi\, kHz$.

Taking the Fourier transform of the equations of motion for fluctuation operators and exploiting the identity $\delta a^\dagger(\omega) = \{\delta a(-\omega)\}^\dagger$, we obtain

$$\delta F_{RP}(\omega_m) = i\hbar g_{OM}^2 a_S^2 \delta x(\omega_m) R \quad (20)$$

where, $R = 1/(S - i\omega_m) - 1/(S^* - i\omega_m)$,

$$S = i\Delta - \frac{J^2}{k_C + i\Delta_C + \frac{g_{at}^2 N}{\gamma_{at} + i\Delta_{at}}} + k_A$$

Due to dynamical back-action induced by the radiation pressure force, the spring constant (thereby the effective oscillation frequency) and the damping rate of the mechanical oscillator get modified. The former is called optical spring effect. The changes in the spring constant $K_{OM}$ and damping constant $\gamma_{OM}$ in terms of the radiation pressure force, are given by [37]:

$$\delta F_{RP}(\omega_m) = -K_{OM}\delta x(\omega_m) + i\omega_m m \gamma_{OM} \delta x(\omega_m) \quad (21)$$

For our system we can identify the optomechanical modification to the mechanical damping rate and spring constant as,

$$\gamma_{OM} = \frac{\hbar(g_{OM}a_S)^2}{m\omega_m} Re(R) \quad (22)$$

$$K_{OM} = \hbar(g_{OM}a_S)^2 Im(R) \quad (23)$$

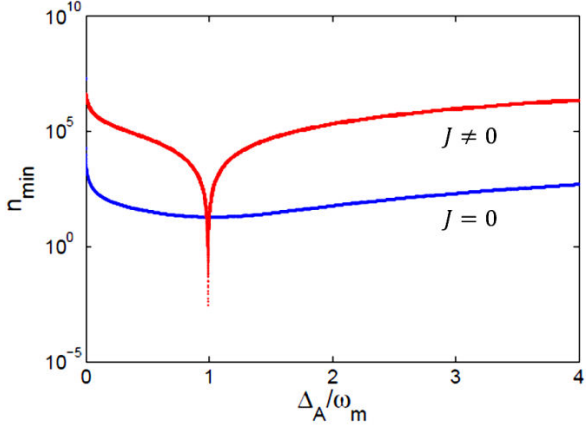

Fig. 6: Plot of $n_{min}$ vs normalized opomechanical cavity detuning $\Delta_A/\omega_m$ for $J = 0$ (blue solid line) and $J \neq 0$ (red solid line), in good cavity limit with $k_A = 2\pi \times 250\ kHz$. Other parameters used are same as Fig. 5.

The minimum number of phonons in the mechanical oscillator is given by:

$$n_{min} = \frac{\gamma_{Stokes} + \gamma_m n_{bath}}{\gamma_{OM} + \gamma_m} \quad (24)$$

where, $n_{bath}$ is the number of bath phonons given by $n_{bath} = \frac{k_B T_{bath}}{\hbar \omega_m}$, with bath temperature given by $T_{bath}$ and $k_B$ being the Boltzmann constant. $T_{bath} \gg 1$ at room temperature.

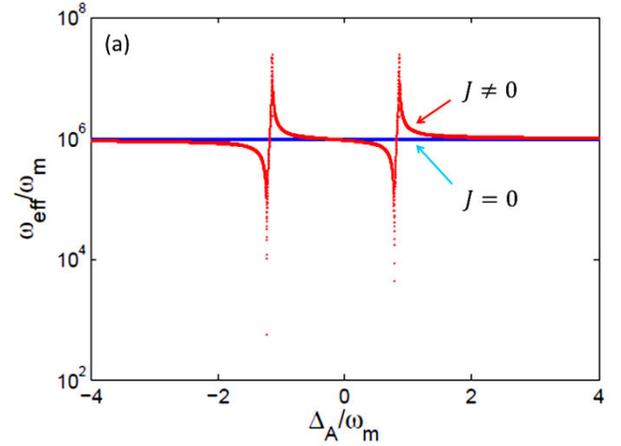

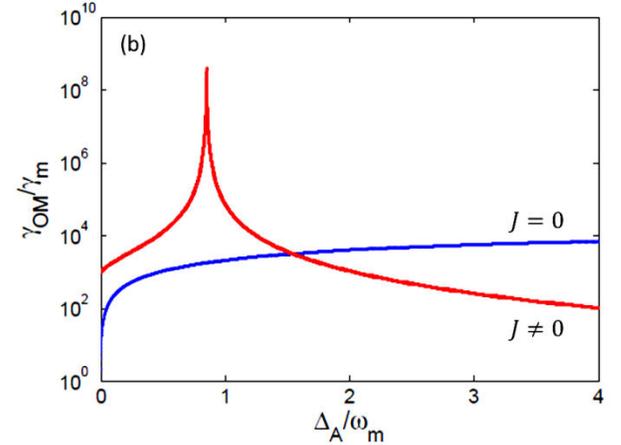

Fig. 7: Plot of (a) $\frac{\omega_{eff}}{\omega_m}$ and (b) $\frac{\gamma_{OM}}{\gamma_m}$ vs $\Delta_A/\omega_m$ in the bad cavity limit for $J = 0$ (blue solid line) and $J \neq 0$ (red solid line), for $P = 3\ \mu W$. Other system parameters used are $L = 1mm$, $m = 10\ ng$, $\lambda = 794.98\ nm$, $k_C = 2\pi \times 21.5\ MHz$, $k_A = 2\pi \times 4\ MHz$, $\omega_m = 2\pi \times 350\ kHz$, $Q = 10^7$, $N = 10^8$, $\gamma_{at} = 2\pi \times 2.875\ MHz$, $\Delta_{at} = 500\gamma_{at}$, $g_{at} = 2\pi\ kHz$.

$\gamma_{Stokes}$ and $\gamma_{anti-Stokes}$ are the two sideband dampings given by

$$\gamma_{Stokes} = \frac{\hbar(g_{OM}a_S)^2}{m\omega_m} Re\left(\frac{1}{S^* - i\,\omega_m}\right)$$

$$\gamma_{anti-Stokes} = \frac{\hbar(g_{OM}a_S)^2}{m\omega_m} Re\left(\frac{1}{S - i\,\omega_m}\right) \quad (25)$$

In Figure 5, we show the variation of $\frac{\omega_{eff}}{\omega_m}$ and $\frac{\gamma_{OM}}{\gamma_m}$ with respect to normalized cavity detuning in absence and presence of the feedback cavity in the good cavity limit, i.e. $k_A \ll \omega_m$. In case of high frequency mechanical oscillators with $\omega_m \geq 1\,MHz$, optical spring effect is not prominent in a general optomechanical setup [38]. But it is interesting to observe that in the hybrid setup, the change in $\omega_{eff}$ is much larger than that in a single cavity setup. Moreover, from Fig. 5(b) we observe that at $\Delta_A = \omega_m$, the optomechanical contribution to the mirror damping for the coupled setup increases nearly by on the order of four compared to that of the uncoupled case. This increase in damping of the mechanical mirror owing to the atomic feedback is the underlying cause of better cooling property revealed in Fig. 6.

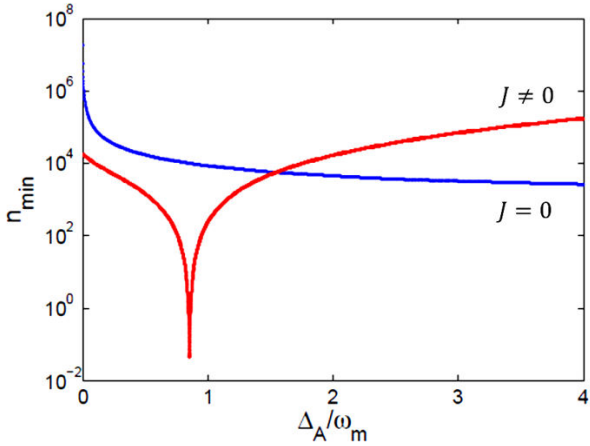

Fig. 8: Plot of $n_{min}$ vs normalized opomechanical cavity detuning $\Delta_A/\omega_m$ for $J = 0$ (blue solid line) and $J \neq 0$ (red solid line), in bad cavity limit with $k_A = 2\pi \times 4\,MHz$. Other parameters used are same as Fig. 7.

In the bad cavity limit also, i.e. for $k_A < \omega_m$, the effective mechanical frequency is highly modulated in presence of the feedback cavity, as observed in Fig. 7. The optomechanical damping is also enhanced, but now for a detuning, $0 < \Delta_A < \omega_m$. Consequently, we have the best cooling in the bad cavity limit for the corresponding cavity detuning, as depicted in Fig. 8. However, this also provides large improvement over the generic cavity cooling schemes that operates mostly in the resolved sideband limit. Therefore, we remark that the hybrid setup is efficient for cooling in both good and bad cavity regimes.

## V. CONCLUSION

In conclusion, we have studied a hybrid system consisting of an optomechanical cavity and another cavity containing ultracold two level atomic ensemble serving as a feedback to the first cavity; with a special emphasize on bistability shown by the cavity optical field and cooling of the mechanical oscillator emerging due to the effect of radiation pressure force. It turns out that the bistable behavior can be controlled by tuning the power and frequency of the single driving laser as well as by changing the atom-cavity coupling in the feedback cavity. This allows more flexibility in controlling bistability compared to the basic single cavity optomechanical system, thus finding the possibility to act as a better optical switch. Furthermore, we have shown that the optomechanical damping is highly enhanced in presence of the coupling and hence the cooling of the mechanical oscillator occurs to be better by more than four orders of magnitude in both good and bad cavity limits.